\documentclass[12pt]{article}
\usepackage{graphicx}
\begin{document}
\begin{center}
{\large\bf Modified Gravitational Theory and the Pioneer 10 and
11 Spacecraft Anomalous Acceleration} \vskip 0.3 true in {\large
J. W. Moffat} \vskip 0.3 true in {\it The Perimeter Institute
for Theoretical Physics, Waterloo, Ontario, N2J 2W9, Canada}
\vskip 0.3 true in and \vskip 0.3 true in {\it Department of
Physics, University of Toronto, Toronto, Ontario M5S 1A7,
Canada} \end{center}
\begin{abstract}%
The nonsymmetric gravitational theory leads to a
modified acceleration law that can at intermediate
distance ranges account for the anomalous acceleration experienced by the
Pioneer 10 and 11 spacecraft.
\end{abstract}
\vskip 0.2 true in
e-mail: jmoffat@perimeterinstitute.ca


\section{Introduction}

A gravitational theory explanation of the acceleration of the
expansion of the universe~\cite{Perlmutter,Riess,Spergel} and the
observed flat rotation curves of galaxies was
proposed~\cite{Moffat,Moffat2}, based on the nonsymmetric gravitational
theory (NGT)~\cite{Moffat3,Moffat4,Moffat5}. The equations of
motion of a test particle in a static, spherically symmetric
gravitational field modifies Newton's law of acceleration for
weak fields. We shall show that the modified acceleration can
account for the anomalous acceleration observed for the Pioneer
10 and 11 spacecraft~\cite{Anderson,Anderson2}. Analyses of
radio Doppler and ranging data from in the solar system indicate
an apparent anomalous acceleration acting on the spacecraft, with a
magnitude $a_P=(8.74\pm 1.33)\times 10^{-8}\,{\rm cm}\,{\rm
s}^{-2}$, directed toward the Sun.

A detailed investigation of both external and internal effects
have not to-date disclosed a convincing source of the anomalous
acceleration. Possible systematic origins of the residuals have
not been found. The gravitational source of the spacecraft
anomalous acceleration can be linked to the increase in the
rotational velocity of a test particle in a circular orbit about
the center of a spiral galaxy. They both may share the same
gravitational origin in the modified acceleration law in NGT.
In the NGT phenomenology, we have suggested that in certain
distance regimes gravity increases its strength with the distance from a
mass source, corresponding to a gravitational ``confinement'' potential.
The gravitational constant is ``renormalized'' by defining the
gravitational constant at infinity to be
\begin{equation}
\label{renormG}
G\equiv G_{\infty}=G_0\biggl(1+\sqrt{\frac{M_0}{M}}\biggr),
\end{equation}
where $G$ is Newton's gravitational constant and $M_0$ is a positive
parameter, which is range-distance dependent, so that as $M_0$ increases
with distance the gravitational constant $G$ increases for a fixed value
of $M$. Conversely, for fixed values of $M_0$ the gravitational constant
$G\rightarrow G_0$ as $M\rightarrow\infty$.

\section{The Equations of Motion of a Particle}

We shall consider the equation of motion~\cite{Moffat2}
\begin{equation} \label{geodesic}
\frac{du^\mu}{d\tau}+\left\{{\mu\atop\alpha\beta}\right\}u^\alpha
u^\beta=s^{(\mu\sigma)}f_{[\sigma\nu]}u^\nu,
\end{equation}
where $u^\mu=dx^\mu/d\tau$ is the 4-velocity of a particle and
$\tau$ is the proper time along the path of the particle.
Moreover, $s^{(\mu\alpha)}g_{(\nu\alpha)}={\delta^\mu}_\nu$ and
\begin{equation}
\left\{{\mu\atop\alpha\beta}\right\}=
\frac{1}{2}s^{(\mu\sigma)}(\partial_\beta g_{(\alpha\sigma)}
+\partial_\alpha g_{(\beta\sigma)}-\partial_\sigma g_{(\alpha\beta)}).
\end{equation}
The notation is the same as in
refs.~\cite{Moffat,Moffat2,Moffat4,Moffat5}. In~\cite{Moffat2}, we
obtained the acceleration on a point particle
\begin{equation}
\label{accelerationlaw}
a(r)=-\frac{G_{\infty}M}{r^2}+\sigma\frac{\exp(-r/r_0)}{r^2}
\biggl(1+\frac{r}{r_0}\biggr),
\end{equation}
where $r_0=1/\mu$ and $\mu$ is the mass parameter in the NGT
action associated with the skew field $g_{[\mu\nu]}$~\cite{Moffat,Moffat2}.
Moreover, $G_{\infty}$ defined in Eq.(\ref{renormG}) is the
effective gravitational constant at infinity where $G_0$ is
Newton's gravitational constant, $G_0=6.673\times 10^{-8}\,{\rm
g}^{-1}\,{\rm cm}^3\,{\rm s}^{-2}$ and $M_0$ and $r_0$ are two free
parameters. The constant $\sigma$ in
(\ref{accelerationlaw}) is given by~\cite{Moffat2}
\begin{equation}
\label{sigma2}
\sigma=\frac{\lambda s G_0^2M^2}{3c^2r_0^2},
\end{equation}
where $\lambda$ and $s$ denote the strengths of the coupling of a
test particle to the force $f_{[\mu\nu]}$ and of the skew field
$g_{[\mu\nu]}$, respectively.

We set $\sigma=G_0\sqrt{M}\sqrt{M_0}$ and write
Eq.(\ref{accelerationlaw}) as
\begin{equation}
\label{NGTaccelerate}
a(r)=-\frac{G_0M}{r^2}\biggl\{1+\sqrt{\frac{M_0}{M}}\biggl[1-\exp(-r/r_0)
\biggl(1+\frac{r}{r_0}\biggr)\biggr]\biggr\}.
\end{equation}
The form of $a(r)$ guarantees that it reduces to the Newtonian acceleration
$a_{\rm Newt}=-G_0M/r^2$ at small distances $r\ll r_0$.

The two parameters $M_0$ and $r_0$ in the NGT solution
for weak fields are not universal constants. The effective
gravitational constant $G=G_{\infty}$ given by (\ref{renormG})
indicates that the strength of gravity is distance dependent and
scales as $\sqrt{M_0/M}$. Moreover, we postulate that the
parameter $M_0$ is dependent on the distance range $r_0$ such
that $M_0=M_0(r_0)$ increases from a small value $M_0 <
M_{\oplus}$ to a large galaxy size mass as the range parameter $r_0$
increases to larger values. We obtained from a fitting of the
galaxy rotation curves $(r_0)_g=13.92\,{\rm kpc}$ and
$(M_0)_g=9.6\times 10^{11}\,M_{\odot}$. For galaxies and clusters of
galaxies the range parameter $r_0$ was determined by
\begin{equation}
a_0=\frac{G_0M_0}{r_0^2}=cH_0,
\end{equation}
where $H_0$ is the current value of Hubble's constant and $cH_0=6.9\times
10^{-8}\,{\rm cm}\,{\rm s}^{-2}$.

The distance-range dependence of the
strength of the gravitational force does not directly come from
the NGT action or the field equations in the weak field
approximation. However, a distance-range dependence could be
built into the action by including a scalar field $\phi=r_0$, so
that $\mu=\mu(\phi)$. It is hoped that an exact solution of the
NGT field equations will reveal the ``running'' of the
gravitational coupling strength with the distance range $r_0$. Another possibility is that
nonlinear, nonperturbative solutions of the field equations lead to
range dependent gravity solutions that deviate significantly from general
relativity for small $g_{[\mu\nu]}$. In the NGT phenomenology, we shall
treat the range parameter $r_0$  and the parameter  $M_0$ as constants
within their approximate valid ranges.

\section{Spacecraft Anomalous Acceleration}

Let us expand the exponental in (\ref{NGTaccelerate}) for
$r<r_0$:
\begin{equation}
\exp(-r/r_0)\biggl(1+\frac{r}{r_0}\biggr)=1-\frac{1}{2}\biggl(\frac{r}{r_0}\biggr)^2
+\frac{1}{3}\biggl(\frac{r}{r_0}\biggr)^3-...
\end{equation} Inserting this into (\ref{NGTaccelerate}) gives for the
magnitude of the anomalous acceleration
\begin{equation}
a_A\equiv a-a_{\rm Newt}
=\frac{G_0\sqrt{M}\sqrt{M_0}}{2r_0^2}.
\end{equation}
For the Pioneer spacecraft we have
\begin{equation}
a_P=\frac{G_0\sqrt{M_{\odot}}\sqrt{(M_0)_{sc}}}{2(r_0)_{sc}^2}.
\end{equation}
By adopting the values $(r_0)_{sc}=300\,{\rm A.U.}=4.49\times
10^{10}\,{\rm km}$, $\sqrt{(M_0)_{sc}}=0.026\sqrt{M_{\odot}}$,
we obtain the Pioneer spacecraft anomalous acceleration~\cite{Anderson2}:
\begin{equation}
a_P=8.7\times 10^{-8}\,{\rm cm}\,{\rm s}^{-2}.
\end{equation}

The anomalous acceleration has not been detected in planetary
orbits, particularly for Earth and
Mars~\cite{Reasenberg,Anderson3}. A small anomalous acceleration
experienced by a planet would cause a perturbed radial
difference $\Delta r$ in the planet's orbit. The anomalous
acceleration $a_A$ would be for a circular orbit
~\cite{Anderson2}
\begin{equation}
a_A =\frac{a_{\rm Newt}\Delta r}{r},
\end{equation}
where $r$ is the orbital radius. For Earth $\Delta r <
-21\,{\rm km}$ and for the mean Sun-Earth radial distance
$r=1\,{\rm A.U.}=1.496\times 10^{8}\,{\rm km}$, we find for
Earth that
\begin{equation}
\label{anomalous}
a_A < 8.32\times 10^{-9}\,{\rm cm}\,{s}^{-2}.
\end{equation}

For the solar planetary system, we adopt the parameter values
\begin{equation}
(r_0)_{pl}\leq 5\times 10^3\,{\rm km},\quad
\sqrt{(M_0)_{pl}} \leq 10^{-19}\,\sqrt{M_{\odot}}.
\end{equation}
For planetary distances, the exponential term in Eq.(\ref{NGTaccelerate})
is vanishingly small, because $r_{pl}\gg (r_0)_{pl}$ and
$\sqrt{(M_0)_{pl}/M_{\odot}}\sim 10^{-19}$ whereby $G\equiv
G_{\infty}\sim G_0$. Thus, any anomalous acceleration would satisfy
$a_A < 3\times 10^{-11}\,{\rm cm}\,{\rm s}^{-2}$ and be undetectable for
solar system planets.

The perihelion advance of Mercury is given by~\cite{Moffat2}:
\begin{equation}
\Delta\omega=\frac{6\pi}{c^2p}(GM_{\odot}-c^2K_{\odot}),
\end{equation}
where $p=a(1-e^2)$ and $a$ and $e$ denote the semimajor axis and
eccentricity of Mercury's orbit, respectively. The parameter
$K_{\odot}$ is
\begin{equation}
K_{\odot}=\frac{\lambda sG_0^2M^2_{\odot}}{3c^4(r_0)^2_{pl}}.
\end{equation}
We require that $K_{\odot} < G_0M_{\odot}/c^2=1.5\,{\rm km}$ to
agree with the observed perihelion advance of Mercury. This
yields the bound
\begin{equation}
\lambda s < \frac{3c^2(r_0)^2_{pl}}{G_0M_{\odot}}
= 5.1\times 10^7\,{\rm km}.
\end{equation} For $(r_0)_{pl}=10\,{\rm km}$ we obtain $\lambda s <
203$ km. The deflection of light grazing the limb of the Sun is given
by~\cite{Moffat2}:
\begin{equation}
\Delta=\frac{4G_0M_{\odot}}{c^2R_{\odot}}
\end{equation}
in agreement with GR.

\section{Conclusions}

We have shown that the modified Newtonian acceleration obtained from NGT
can with a suitable choice of the two parameters $r_0$ and $M_0$ explain
the anomalous acceleration observed for the Pioneer spacecraft 10 and 11.
The result depends on the phenomenological premise that the strength of
gravity is distance dependent. The range parameter has a growing dependence
on distance as does the parameter $M_0$. The running of the gravitational
constant $G$ with mass as $\sqrt{M_0/M}$ is a fundamental physical
phenomenon that could have its origins in exact solutions of the NGT field
equations. This issue will be the subject of a future investigation.

A possible signature of the NGT modified Newtonian acceleration, and the
anomalous acceleration for spacecraft moving away from the Sun, is to
observe whether there is an eventual increase in the anomalous acceleration
as the spacecraft moves to distances from the Sun greater than, say, 200 or
300 A.U., revealing a range dependence on the parameter $r_0$.

\end{document}